\documentclass[manuscript,nonacm]{acmart}

\AtBeginDocument{%
  }

\newcommand{\percentagepoint}[1]{percentage point~}

\begin{document}

\title[Longitudinal Study on Social and Emotional Use of AI Conversational Agent]{Longitudinal Study on Social and Emotional Use of AI Conversational Agent}

\author{Mohit Chandra}
\email{mchandra9@gatech.edu}
\affiliation{%
  \institution{Georgia Institute of Technology}
  \country{USA}
}

\author{Javier Hernandez}
\email{javierh@microsoft.com}
\affiliation{%
  \institution{Microsoft Research}
  \country{USA}
}

\author{Gonzalo Ramos}
\email{goramos@microsoft.com}
\affiliation{%
  \institution{Microsoft Research}
  \country{USA}
}

\author{Mahsa Ershadi}
\email{mahsaershadi@microsoft.com}
\affiliation{%
  \institution{Microsoft}
  \country{Canada}
}

\author{Ananya Bhattacharjee}
\email{ananya@cs.toronto.edu}
\affiliation{%
  \institution{University of Toronto}
  \country{Canada}
}
\author{Judith Amores}
\email{judithamores@microsoft.com}
\affiliation{%
  \institution{Microsoft Research}
  \country{USA}
}
\author{Ebele Okoli}
\email{ebeleokoli@microsoft.com}
\affiliation{%
  \institution{Microsoft}
  \country{USA}
}
\author{Ann Paradiso}
\email{annpar@microsoft.com}
\affiliation{%
  \institution{Microsoft Research}
  \country{USA}
}

\author{Shahed Warreth}
\email{swarreth@microsoft.com}
\affiliation{%
  \institution{Microsoft}
  \country{Ireland}
}
\author{Jina Suh}
\email{jinsuh@microsoft.com}
\affiliation{%
  \institution{Microsoft Research}
  \country{USA}
}

\renewcommand{\shortauthors}{Chandra et al.}

\begin{abstract}
Development in digital technologies has continuously reshaped how individuals seek and receive social and emotional support. While online platforms and communities have long served this need, the increased integration of general-purpose conversational AI into daily lives has introduced new dynamics in how support is provided and experienced. Existing research has highlighted both benefits (e.g.,~wider access to well-being resources) and potential risks (e.g.,~over-reliance) of using AI for support seeking. In this five-week, exploratory study, we recruited 149 participants divided into two usage groups: a baseline usage group (BU, $n=60$) that used the internet and AI as usual, and an active usage group (AU, $n=89$) encouraged to use one of four commercially available AI tools (Microsoft Copilot, Google Gemini, PI AI, ChatGPT) for social and emotional interactions. Our analysis revealed significant increases in perceived attachment towards AI (32.99 percentage points), perceived AI empathy (25.8 p.p.), and motivation to use AI for entertainment (22.90 p.p.) among the AU group. We also observed that individual differences (e.g., gender identity, prior AI usage) influenced perceptions of AI empathy and attachment. Lastly, the AU group expressed higher comfort in seeking personal help, managing stress, obtaining social support, and talking about health with AI, indicating potential for broader emotional support while highlighting the need for safeguards against problematic usage. Overall, our exploratory findings underscore the importance of developing consumer-facing AI tools that support emotional well-being responsibly, while empowering users to understand the limitations of these tools.
\end{abstract}


\maketitle

\section{Introduction}

The rapid proliferation of digital technologies has fundamentally reshaped how individuals seek and receive social and emotional support. Traditional face-to-face interactions have increasingly been complemented—or even replaced—by online platforms. Social media and online communities have become essential spaces for emotional expression, connection, and support~\cite{mckinseyMentalHealth, naslund2016future}. In recent years, the growing popularity of generative AI-based conversational agents (e.g., OpenAI ChatGPT, Microsoft Copilot, Google Gemini) has driven a surge in individuals using these tools for productivity and well-being support~\cite{wamba2020influence, Polak2024, Li2023, emarketerGenerativeAdoption, newsweekPeopleUsing}. Advances in AI's ability to generate human-like and emotionally intelligent responses have expanded existing well-being support opportunities, leading more people to turn to these tools for their well-being needs~\cite{washingtonaitherapy2024, bbcCharacteraiYoung, alanezi2024assessing, song2024typing, siddals2024happened, ma2024understanding, marriott2024one}.

As AI becomes increasingly integrated into daily life, it is crucial to examine its broader implications for users' perceptions on AI that may impact their mental health and societal dynamics down the line. Research on AI's impact on emotional and mental well-being has highlighted both significant benefits and potential risks. Previous research has highlighted benefits such as increased access to emotional support and reduced stigma~\cite{song2024typing, alanezi2024assessing, doi:10.1177/20427530241259099}, alongside potential negative consequences such as the development of problematic attachment to AI~\cite{guerreiro2023attracted, pentina2023exploring}, over-reliance on AI~\cite{jacobs2021machine, chiesurin-etal-2023-dangers}, erosion of user trust in AI~\cite{doi:10.5465/annals.2018.0057, chiesurin-etal-2023-dangers, shah2022goal}, and reduced human connection \cite{davenport2020artificial, grewal2024ai}. While these studies have enhanced our understanding of specific aspects of human-AI interaction, a comprehensive view of the holistic impact on users' emotional and psychological well-being remains under-explored. Furthermore, much of the existing research has relied on cross-sectional methods that capture only immediate effects, while longitudinal studies have often depended on retrospective surveys or interviews~\cite{skjuve2021my, skjuve2022longitudinal, pentina2023exploring}. This leaves a gap in understanding how sustained interactions with AI might shape long-term user perceptions and usage motivations.

\begin{figure}[t]
    \centering
    \includegraphics[width=\linewidth]{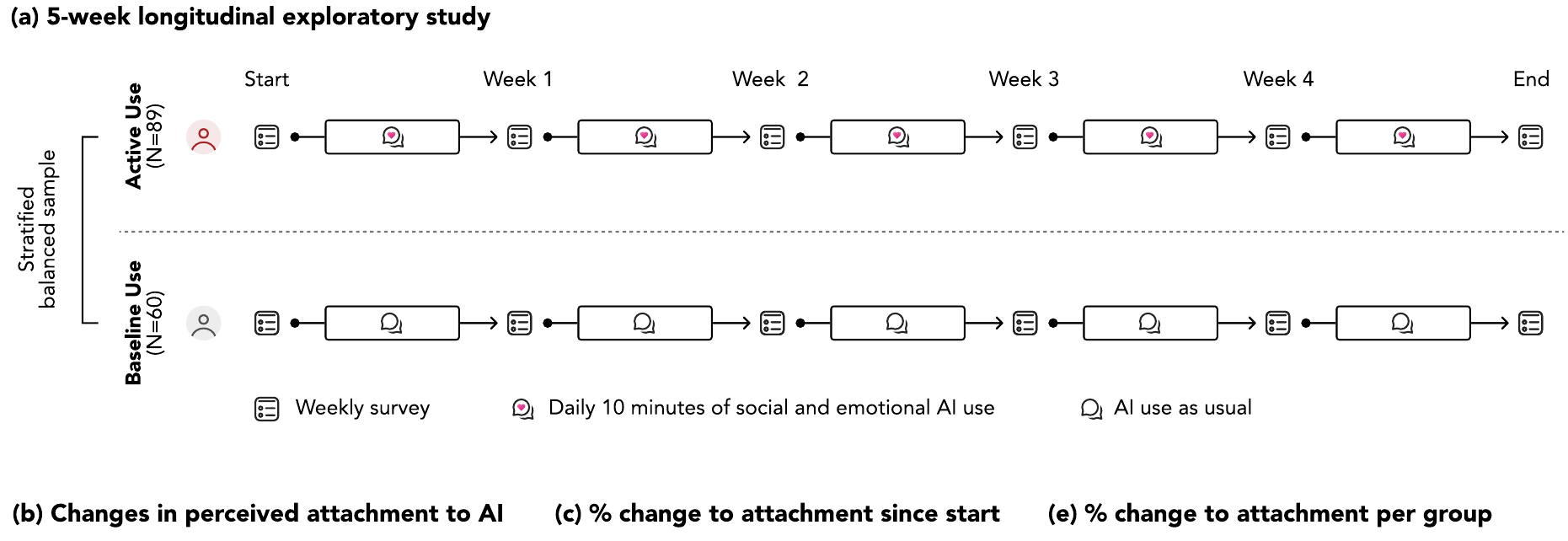}
    \includegraphics[width=\linewidth]{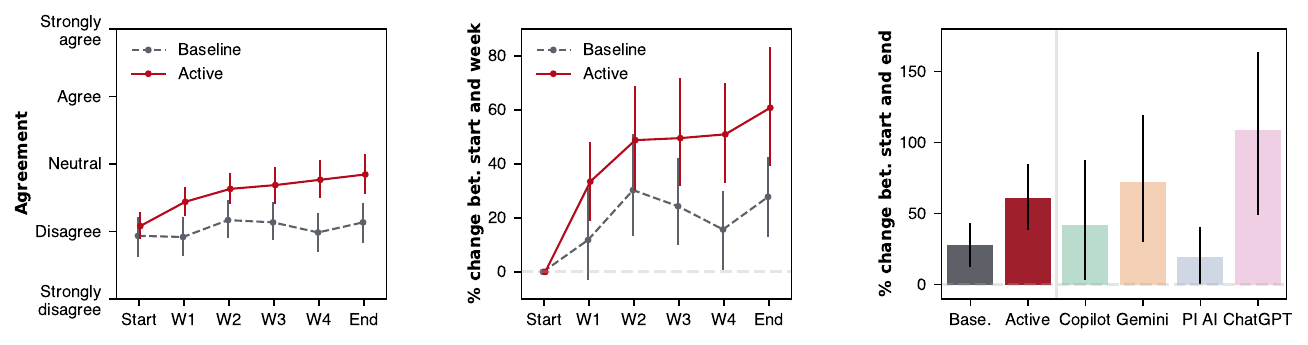}
    \caption{\textbf{Studying the impact of social and emotional use of generative conversational AI agents on perceived attachment to AI.}
    \textbf{(a)}~Participants are divided into baseline usage (BU) or active usage (AU) groups, where each group retains the equivalent balance distribution. AU participants were asked to use their assigned AI for social and emotional scenarios each day for at least 10 minutes. BU participants were asked to use AI as usual. All participants were asked to complete weekly surveys from the start to end of the study.
\textbf{(b)}~Weekly average self-reported agreement to the statement ``I feel attached to [assigned AI conversational agent]'' on a 5-point Likert scale.
\textbf{(c)}~Average percentage change at each subsequent week compared to the study start. 
\textbf{(d)}~Average percentage change at the end of the study compared to the study start for each subgroup. 
In \textbf{(b)}-\textbf{(d)}, data are presented as mean values, and error bars indicate 95\% confidence intervals.}
    \label{fig:figure1_attachment}
\end{figure}

In this five-week longitudinal exploratory study, we aimed to examine how interactions with commercially available AI agents influence user experiences and perceptions over time. We recruited 149 participants divided into two usage groups: a baseline usage group (BU, $n=60$) that continued their typical internet and AI usage, and an active usage group (AU, $n=89$) assigned to use one of four commercially available AI platforms: OpenAI ChatGPT~\cite{achiam2023gpt}, Microsoft Copilot~\cite{microsoftMicrosoftCopilot}, Google Gemini~\cite{googleGeminiChat}, and PI AI~\cite{piai} for social and emotional interactions (e.g., discussing personal struggles, building emotional connections with AI). We balanced both groups across variables such as gender identity, AI usage tendency, self-reported presence of a mental health condition (Fig.~\ref{fig:figure1_attachment}a, Methods). Using weekly surveys, we measured participants' self-reported perceptions, such as motivation for using AI~\cite{huang2024ai}, perceived empathy in AI~\cite{schmidmaier2024perceived}, dependence on these tools~\cite{huang2024ai}, and changes in interpersonal orientation~\cite{filsinger1981measure} (Fig.~\ref{fig:figure1_attachment}b-c, Methods). 
We then employed the difference-in-difference (DiD) approach~\cite{lechner2011estimation} to estimate how active social and emotional interactions influenced these perceptions (Fig.~\ref{fig:figure1_attachment}d-e, Methods). Additionally, we examined how participant characteristics modulated their evolving perceptions of AI. To complement these findings, we collected and analyzed participants' open-ended remarks on their behavior and perception shifts, as well as their design recommendations for improving the safety and usefulness of AI conversational agents.

\section{Methods}

\subsection{Recruitment and Participants}

Our recruitment strategy focused on ensuring a balanced set of participants between active usage (AU) and baseline usage (BU) groups and among four platform groups (Copilot, Gemini, PI AI, ChatGPT).
For this, we balanced the groups for three potentially confounding variables captured in the screening survey -- gender identity, AI usage tendency, and self-reported presence of a mental health condition. Our screening survey included several parts. First, we described the study details and asked the participants to confirm their commitment to adhering to the study protocol and indicate whether they required any accessibility accommodations. Additionally, we collected demographic information, such as age, gender identity, highest education attainment, race/ethnicity, household size and income, and self-reported presence of a mental health condition. Next, to assess participants’ AI usage tendency tendencies, we collected the frequency of the usage of the four AI platforms. Finally, to examine the traits linked to problematic internet and technology use, we also incorporated previously validated scales for assesing self-reported impulsivity (I-8~\cite{groskurth2022impulsive}), loneliness (Three-Item Loneliness Scale~\cite{hughes2004short}), and locus of control (Internal Control Index~\cite{duttweiler1984internal}). These constructs have been associated with problematic internet or technology use in past research~\cite{li2021relationship, hu2023social, epley2008creating, chak2004shyness}.

Participant distribution across groups was balanced using stratified sampling (Figure~\ref{fig:figure1_attachment}a). We balanced on self-reported gender identity because gender was found to influence emotional disclosure~\cite{de2017gender}, expression~\cite{wolf2000emotional}, and support-seeking behaviors~\cite{tifferet2020gender}. We did not have enough interested participants who identified as non-binary to include them equally across all groups. Past research has shown that individuals with mental health conditions often engage with online peer support~\cite{naslund2016future} or online mental health services~\cite{kauer2014online} as alternatives or supplements to in-person counseling. To control for the potential bias introduced by people's preferences or past experiences towards technology-facilitated emotional support, we balanced the groups on self-reported presence of a mental health condition. For gender identity and mental health conditions, we tried to optimize for 50/50 balance in each binary category. The extent to which AI is integrated into the daily life of an individuals has been shown to influence human-AI relationships~\cite{skjuve2022longitudinal}. To account for this, after categorizing potential participants into buckets of ``hot'' users (i.e., use one or more of the four specified AI platforms more than once a week) and the rest of the users as ``cold'', we balanced for AI usage by randomly distributing participants to maximize ``cold'' users and minimize ``hot'' AI usage tendency users. We also skewed participant sampling towards high loneliness equally across groups in hopes of a larger effect on social and emotional use of AI. 

We recruited participants through a screening survey conducted on Prolific\footnote{\url{https://www.prolific.com/}}. We used Prolific's filter sets to bias our sampling towards individuals with prior experience with at least one of the four AI platforms and those with self-reported presence of a mental health condition. Our power analysis indicated that we needed 64 participants per group (BU and AU groups) to see medium effect size. Hence, we oversampled in anticipation of a heavier than normal dropout, which we discuss next. In total, 547 participants completed the screening survey and were interested in participating in the study, and 240 were officially invited to participate in the study after balancing. However, we faced a significant challenge with participant retention due to the compensation design. We anticipated a high attrition rate percentage due to the deviation of our study from typical Prolific tasks in both duration and complexity of the study. While Prolific provides guidelines for setting up longitudinal or multi-part studies\footnote{\url{https://researcher-help.prolific.com/en/article/620ca2}},  the platform's compensation structure necessitates separate payments for each study component. 
While providing a higher bonus reward at the completion of the study by potentially reducing the per-task hourly rate is recommended to encourage retention, we prioritized for more ethical compensation practices (e.g., see \cite{thurman2015ethical} for the precarious conditions of crowd workers) and paid them the recommended hourly rate per each part of the study regardless of their study completion status. This compensation approach eliminated financial penalties for attrition but potentially increased dropout rates. Although we had a 25.13\% dropout overall, we were able to maintain the sampling balance we intended.

In total, 149 participants completed the study: 60 in BU, 23 in ChatGPT, 22 in PI AI, 23 in Copilot, 21 in Gemini. Chi-square test of independence ($\tilde{\chi}^2$) revealed no significant associations between groups (AU vs. BU, between the four Platform groups, or the four Platform subgroups and BU group) and categorical variables such as gender identity, self-reported presence of a mental health condition, or AI usage tendency.  For numerical variables, no significant differences were observed between the groups for self-reported Locus of Control, Loneliness, or Impulsivity scores. Regarding gender identity, 53.9\% of participants in the AU group identified as male, compared to 48.3\% in the BU group.~57.3\% of participants in the AU group self-reported having a mental health condition, compared to 48.3\% in the BU group. Additionally, for the AI usage tendency, a larger proportion of participants were classified as ``cold users,'' (participants using AI less than once a week) with 56.2\% in the AU group and 61.7\% in the BU group. For Locus of Control, AU participants had a mean score of 100.23 ($\sigma$=15.95), whereas the BU group had a mean score of 100.85 ($\sigma$=13.65). Loneliness score was distributed between 4 to 9 for both groups, with participants in the AU group having a mean score of 6.56 ($\sigma$=1.59) and a mean score of 6.27 ($\sigma$=1.49) for the BU group. Finally, the impulsivity score for AU group participants averaged at 20.17 ($\sigma$=5.51) and 21.75 ($\sigma$=5.24) for the BU group participants. Supplementary Figures 1-4 present participant descriptive statistics across these variables in all groups.

\subsection{Study Procedure}
Our 5-week exploratory study aimed to examine between-subject differences in perception changes, primarily between the AU and BU groups, and secondarily across four Platform groups (Figure~\ref{fig:figure1_attachment}b). The BU group was asked to use the internet, AI agents, or other technology tools as they normally would. This group served as a baseline, controlling for potential seasonal or external factors that could influence AI usage tendencies and perceptions. BU participants were compensated with a \$50 gift card for full participation of approximately 3 hours and 15 minutes. AU participants were each assigned to a specific AI agent out of the four platforms and were asked to incorporate the assigned AI conversational agent into their daily lives as they normally would and to use the agent for at least 10 minutes a day on one or more of the following emotional and social scenarios: (1) Talk through struggles in or reflect on your life or close relationships, (2) Talk through job, school, or finance-related stress, anxiety, burnout, (3) Ask for help navigating new or challenging social situations, (4) Try to build friendships or deep emotional relationships with the AI agent, or (5) Talk through physical and psychological symptoms or discomforts. While the AU participants were allowed to use other AI agents for work, school, or other tasks, we encouraged them to use the assigned AI agent for the above five scenarios and to refrain from using any other AI agents or services as much as possible. We provided each AU participant with a user account dedicated to the study that they can use to log in and interact with the assigned AI agent. Each account was pre-loaded with a two-month subscription to the highest level of pricing tier available for the platform (i.e., Copilot PRO, ChatGPT Plus, Gemini Advanced; PI AI did not have a pricing tier). 

To ensure participants' adherence to study protocols and to protect participant privacy, we implemented reminder protocol and conducted weekly data quality checks. Specifically, we provided multiple reminders to the participants at each interaction point that research personnel had access to all AI conversations for quality assurance and data collection purposes. These reminders explicitly instructed participants to avoid disclosing personally identifiable information during their AI interactions, supporting both ethical research practices and data quality. To minimize platform-specific confounds, we standardized the interaction medium by requiring participants to use only the web-based versions of the AI agents. Furthermore, participants were instructed to engage through text-based interactions rather than utilizing multimodal features (e.g., voice, image, or video inputs) that vary across platforms. AU participants were compensated with a \$115 gift card for full participation of approximately 7 hours and 20 minutes. This study was conducted in July and August of 2024. The study protocol was reviewed and approved by the institution's ethics review board.

\subsubsection{Weekly Measures}

Our exploratory study was structured to capture a total of six weekly snapshots of participant perception along the perception variables. These snapshots were taken at the intake survey at the beginning of Week 1, four weekly surveys, and an exit survey after Week 5. Additionally, all surveys shared a common set of instruments that capture perception variables on a weekly basis. 

The common set of instruments included the following scales: (1) Motivation for using AI (AI Use Motivation Scale~\cite{huang2024ai}) along escape, social, entertainment, and instrumental subscales, (2) percieved problematic dependence on AI (AI Dependence~\cite{huang2024ai}) that captured the percieved reliance, excessive use, jeopardization, withdrawal, and loss of control symptoms, (3) percieved interpersonal orientation (Liking People Scale~\cite{filsinger1981measure}) that measured how much did the participant liked spending time with others or are socially anxious, (4) attitude towards AI (General Attitudes towards Artificial Intelligence Scale~\cite{schepman2020initial} with positive and negative subscales, and (5) perceived empathy of AI (Perceived Empathy of Technology Scale (PETS)~\cite{schmidmaier2024perceived}) with emotional responsiveness and understanding and trust subscales. We did not include two questions from the PETS questionnaire as those did not focus on the emotional aspects of perceived empathy from the AI system.

To assess participants' AI usage tendency, we collected data on the total duration and frequency of interactions with both the assigned AI and any other AI agents. In addition to capturing the perceived empathy and general attitude towards AI, we also included other perception metrics, such as their percieved attachment towards AI, percieved  satisfaction with AI, percieved helpfulness of AI, percieved human-like behavior of AI, and whether or not they would recommend AI to others. Lastly, we also included open-ended questions around the purpose of AI use, whether the participants' purpose of use of AI has changed, general impressions about AI, positive or negative impacts from using AI, any problematic experiences with using AI, and any suggestions for AI designers. When referring to AI, BU participants were asked to comment on any AI agent in general while AU participants were asked to comment on their assigned AI agent. Below, we describe participant activities in each week, highlighting any per-group or weekly differences. We have provided the list of questions in the Supplementary section (Questionnaire for Custom Questions)

\subsubsection{Week 1 - Intake}

At the start of Week 1, participants received an intake survey containing the study information sheet and a set of initial screening questions. The study information was customized to ensure participants provided explicit consent for conducting per-group activities. After providing the consent, we presented the common set of questions described above and a set of open-ended questions probing participants towards initial motivation for using AI. In addition, AU participants were presented with a familiarity task aimed at ensuring that they were able to access their assigned AI and to gain familiarity with the platform's capabilities and the study requests (i.e., interact 10 minutes a day for social and emotional scenarios). For this task, we provided the login instructions and asked the AU participants to verify access to the account and converse with the assigned AI agent along 8 different scenarios. These 8 scenarios included five social and emotional scenarios we described above as well as 3 non-emotional scenarios, such as (1) Get help with learning a new skill, (2) Get help with writing or drafting, and (3) Get help with coming up with new ideas for work-related aspects or for planning events. For each of these scenarios, we provided participants with a suite of sample prompts to choose from. We performed random checks on 20\% of accounts and followed up with participants who marked the task as complete but did not follow the instructions.

\subsubsection{Weeks 2 to Week 5 - Weekly}

At the beginning of each week, we released a weekly survey containing all of the common weekly measures described above. For AU participants, we provided additional reminders to use their assigned AI for 10 minutes daily for requested scenarios. For the duration of the study, we conducted random checks for adherence each week for roughly 10\% of the participants and communicated with non-compliant participants to get them back on track. 

\subsubsection{After Week 5}
\label{sec:methods_week_5}

At the conclusion of Week 5, all of the participants were provided with an exit survey that contained the common weekly measures as well as several exit-only questions. We asked participants to rate their perception of the usefulness of AI for the 8 scenarios used in the familiarity task, including whether they have or have not used AI for those scenarios. We also asked participants how their comfort level in using AI for each of the five social or emotional support scenarios had changed, and asked them to explain why those changes occurred. To gain additional nuances in percpetion towards AI that were not explicitly covered by our weekly measures, the exit survey contained questions regarding 10 potential negative outcomes. These negative outcome questions included: (1) My mental health issues were exacerbated by my use of AI, (2) AI is not supportive of my emotional and social needs, (3) I developed negative self-perception and loss of autonomy due to my AI use, (4) I rely on AI for decisions and advice on important aspects of my life, (5) I use AI as a primary source of mental health and emotional support, (6) I use AI as a primary source of social interaction, reducing time spent with friends and family, (7) I developed an emotional attachment towards AI, (8) My use of AI led to social and interpersonal issues with other humans, (9) I have fears about the future related to my reliance on AI, and (10) I believe AI can replace the need for human emotional and social support (Figure~\ref{fig:figure4_comfort}b). For each of these statements, we captured agreement as well as open-ended reasons for their rating. We asked participants to select up to 3 of these outcomes they feel are likely to happen if they were to continue interacting with AI. Finally, to understand future design opportunities, we asked participants to choose the top 3 areas that AI designers should focus or not focus on to reduce user dependency on AI, as well as any open-ended suggestions.

This study was reviewed by the Microsoft Research Institutional Review Board, which is federally registered with the Office for Human Research Protections (OHRP) prior to the research activities. 

\subsection{Analysis}

Our study was designed to answer two main research questions: (1) How does emotional and social use of AI change individual perceptions of AI? and (2) How do individual characteristics contribute to these changes? 
All analyses were done in Python using standard quantitative analysis and visualization libraries, such as pandas, numpy, scipy, statsmodels, and seaborn. 

We analyzed the changes in percpetion variables from the beginning to the end of the study across groups, isolating the effects of the emotional use of AI using the difference-in-difference (DiD) method~\cite{lechner2011estimation}.  We note that there were no statistically significant differences between AU and Control groups in these outcome measures at the start of the study, with the notable exception of perceived satisfaction where the AU group had a lower perceived satisfaction with AI than Control (3.41 vs 3.73 out of a 5-point scale, t=2.68, p$=$0.008) and AI helpfulness where AU had lower perceived helpfulness from AI (3.61 vs 3.95 out of a 5-point scale, t=2.78, p$=$0.006). To assess the normality of within-subject change scores, we conducted Shapiro–Wilk tests for each perception variable within both the AU and Control groups~\cite{shapiro1965analysis}. For the AU group, changes in Attitude towards AI (Positive), Attitude towards AI (Negative), Interpersonal orientation, and Perceived AI empathy were normally distributed. In the BU group, the changes in Attitude towards AI, Attitude towards AI (Negative), and Perceived AI empathy met normality assumptions. Accordingly, we applied paired t-tests to these variables to assess within-subject changes (i.e., change between intake and exit measures) within each group. For perception variables that violated normality assumptions, we used the non-parametric Wilcoxon signed-rank test~\cite{conover1999practical} to evaluate within-subject changes. Table~\ref{tab:relative_change_alltreatment_control} presents DiD in percentage points and relative changes in percentages for AU and Control group comparisons. Supplementary Table 3-6, 8-11 presents the results for Platform group comparisons.  To examine how individual characteristics contribute to changes observed in the perception variables, we modeled the perception variables using a linear mixed effects model. The model incorporated per-participant variables, such as gender identity, self-reported presence of a mental health condition, and AI usage tendency, along with their respective group and interaction terms between individual characteristics and group as fixed effects. Participant ID was included as a random effect. We conducted this analysis between AU and Control groups and among four Platform groups.

\begin{align*}
    \mathsf{Perception~Value}_{\mathsf{diff}} &\sim \mathsf{Group} + 
 \mathsf{Characteristics} + \mathsf{Characteristics} \ast \mathsf{Group} + (1|\mathsf{p}_{\mathsf{id}})\\[-1ex] 
    &\mathsf{where~} \mathsf{Group} \in \{\mathsf{BU, AU, ChatGPT, Copilot, Gemini, PI}\}\\[-1ex] 
    &~\mathsf{and~} \mathsf{p}_{\mathsf{id}} \mathsf{models~the~random~effect~for~each~participant}
\end{align*}

To account for the impact of the limited duration of our study, we conducted a temporal analysis to assess the influence of both time and the AU group on the perception variables. This analysis aimed to ensure the robustness of the observed impact of the active usage among the participants to evaluate the change in participant perceptions after the study. We modeled this effect using a regression model for perception variable $v$ at time $t$.

\begin{align*}
    \mathsf{Perception~Value}_{\mathsf{v}}^{\mathsf{t}} &\sim \mathsf{Time} \ast \mathsf{Group} + 
 \mathsf{Time} + \mathsf{Group} +  \mathsf{PerceptionValue}_{\mathsf{v}}^{\mathsf{Intake}}\\[-1ex] 
    &\mathsf{where~} \mathsf{Group} \in \{\mathsf{BU, AU, ChatGPT, Copilot, Gemini, PI}\}
\end{align*}

We did not observe any significant interaction between AU and time for any perception variable. Compared to the Control group, participants in the AU groups demonstrated higher perceived empathy from AI by 0.52 points (5-point scale; p$<$0.001; t$=$3.88; std$=$0.13; 95\% CI [0.25, 0.78]). We observed a similar impact for satisfaction with AI where AU participants reported 0.58 points higher satisfaction compared to Control (5-point scale; p$<$0.001; std$=$0.16; t$=$3.68; 95\% CI [0.27, 0.89]). Additional results are provided in Supplementary Tables 12-31.

\subsection{Exit survey specific question analysis}

At the end of the study, we asked participants to report perceived changes in their comfort levels for using AI conversational agents for five social and emotional scenarios: seeking personal help, managing stress, obtaining social support, seeking companionship, and talking about health. We measured these changes on a scale of -2 (Much less) to 2 (Much more). To evaluate between-subject differences among the AU and Control group for the comfort changes, we conducted t-test followed by Benjamini-Hochberg procedure to control the False Discovery Rate (FDR) that may arise during multiple hypothesis testing. For comparisons across Platform groups, we performed ANOVA test followed by Tukey’s Honestly Significant Difference (HSD) post hoc test to identify pairwise differences among Platform groups. We also analyzed participant responses towards 10 potential negative outcomes (as described in Section~\ref{sec:methods_week_5}). Similar to the comfort change analysis we conducted t-test followed by Benjamini-Hochberg procedure to assess the between-subject differences among the AU and Control group. For platform groups, we performed ANOVA test followed by Tukey’s Honestly Significant Difference (HSD) post hoc test.

\section{Results}
We present our findings by first, discussing the main differences in perception changes observed between baseline usage (BU) and active usage (AU) groups; second, identifying participant characteristics that had a significant impact on those differences; and third, sharing design recommendations from participants for mitigating potential risks introduced by social and emotional use of AI.

\begin{table}[t]
\centering
\sffamily
\footnotesize
\begin{tabular}{l|ccc|ccc|ccc}
\toprule
\textbf{Perception variable}               & \textbf{DiD (p.p.)} & \textbf{95\% CI}    & \textbf{P value} & \textbf{\begin{tabular}[c]{@{}c@{}}Base. usage \\ \% change\end{tabular}} & \textbf{95\% CI}   & \textbf{P value} & \textbf{\begin{tabular}[c]{@{}c@{}}Active usage \\ \% change\end{tabular}} & \textbf{95\% CI}   & \textbf{P value} \\
\midrule
\multicolumn{10}{l}{\textbf{(a) Motivation for using AI}} \\
\midrule
Motivation for Using AI (Entertainment) & 22.90      & {[}~2.88, 42.92{]}   & 0.025  & 9.17              & {[}-2.14, 20.48{]} & 0.510   & 32.07               & {[}17.31, 46.83{]} & $<$0.001   \\
Motivation for Using AI (Escape)        & 14.89      & {[}-4.95, 34.73{]}  & 0.141 & 9.46              & {[}-4.52, 23.45{]} & 0.700   & 24.35               & {[}10.84, 37.87{]} & 0.028   \\
Motivation for Using AI (Social)        & 17.91      & {[}-3.24, 39.06{]}  & 0.097   & 20.31             & {[}~3.62, 36.99{]}  & 0.050   & 38.22               & {[}24.71, 51.73{]} & $<$0.001   \\
Motivation for Using AI (Instrumental)  & 12.23      & {[}-0.45, 24.91{]}  & 0.059   & 5.33              & {[}~-1.39, 12.06{]}  & 0.311   & 17.56               & {[}~8.07, 27.06{]}  & $<$0.001   \\
\midrule
\multicolumn{10}{l}{\textbf{(b) Perception towards AI}} \\
\midrule
Attachment towards AI                   & 32.99      & {[}~~2.32, 63.66{]}   & 0.035  & 27.78             & {[}12.34, 43.22{]} & 0.062   & 60.77               & {[}37.56, 83.98{]} & $<$0.001   \\
Perceived AI Empathy                    & 25.80      & {[}~10.82, 40.78{]}  & $<$0.001   & 8.93              & {[}~1.01, 16.86{]}  & 0.079   & 34.73               & {[}23.51, 45.95{]} & $<$0.001   \\
Satisfaction with AI                    & 11.25      & {[}~~0.19, 22.32{]}   & 0.046   & 6.61              & {[}~-0.03, 13.26{]}  & 0.067   & 17.87               & {[}~9.85, 25.88{]}  & $<$0.001   \\
AI Helpfulness                          & ~7.74       & {[}~-2.72, 18.20{]}   & 0.146   & 7.17              & {[}~0.28, 14.05{]}  & 0.059   & 14.91               & {[}~7.57, 22.25{]}  & $<$0.001   \\
Perceived Human-Like Behavior           & 14.68      & {[}~-5.98, 35.34{]}  & 0.163   & 8.44              & {[}-1.31, 18.20{]} & 0.389   & 23.13               & {[}~7.31, 38.95{]}  & 0.014  \\
Recommendation to Use AI                & -3.25      & {[}-21.15, 14.66{]} & 0.721  & 21.81             & {[}~3.03, 40.58{]}  & 0.013   & 18.56               & {[}10.63, 26.49{]} & $<$0.001   \\
Attitude towards AI                     & ~1.84       & {[}~~-1.91, ~5.58{]}   & 0.335   & 0.44              & {[}-2.34, ~3.21{]}  & 0.831   & 2.27                & {[}~-0.21, ~4.76{]}  & 0.102   \\
Attitude towards AI (Positive)          & ~3.21       & {[}~~-0.90, ~7.32{]}   & 0.126   & -1.10             & {[}-4.34, ~2.15{]}  & 0.386   & 2.11                & {[}~-0.52, ~4.74{]}  & 0.105   \\
Attitude towards AI (Negative)          & -0.93      & {[}~~-6.85, ~4.98{]}   & 0.756   & 3.46              & {[}-1.96, ~8.88{]}  & 0.330   & 2.53                & {[}~-0.78, ~5.83{]}  & 0.141  \\
\midrule
\multicolumn{10}{l}{\textbf{(c) Dependence on AI}} \\
\midrule
Dependence on AI                        & -0.72      & {[}-11.69, 10.25{]} & 0.897   & 9.68              & {[}-0.26, 19.63{]} & 0.383   & ~8.96                & {[}~2.77, 15.16{]}  & 0.073   \\
Dependence on AI (Over-reliance)        & ~4.98       & {[}-11.69, 21.65{]} & 0.557   & 10.28             & {[}-1.75, 22.30{]} & 0.559   & 15.26               & {[}~4.03, 26.49{]}  & 0.185   \\
Dependence on AI (Excessive Usage)    & -1.11      & {[}-16.15, 13.94{]} & 0.885   & 9.72              & {[}-2.84, 22.29{]} & 0.840   & ~8.61                & {[}-0.60, 17.83{]} & 0.862   \\
Dependence on AI (Jeopardization)       & -0.23      & {[}-10.83, 10.36{]} & 0.965   & 4.17              & {[}-6.81, 15.14{]} & 1.000   & ~3.93                & {[}-0.89, ~8.76{]}  & 0.257   \\
Dependence on AI (Withdrawal)           & ~2.24       & {[}-16.39, 20.87{]} & 0.813   & 16.11             & {[}-0.36, 32.59{]} & 0.221   & 18.35               & {[}~7.53, 29.17{]}  & 0.016   \\
Dependence on AI (Loss of Control)    & -9.34      & {[}-23.11,~ 4.42{]}  & 0.183  & 18.33             & {[}~4.89, 31.78{]}  & 0.040   & ~8.99                & {[}~1.97, 16.01{]}  & 0.071   \\
\midrule
\multicolumn{10}{l}{\textbf{(d) Interpersonal orientation}} \\
\midrule
Interpersonal Orientation               & 3.16       & {[}-0.89, 7.20{]}   & 0.126   & 2.22              & {[}-1.27, 5.70{]}   & 0.119   & 5.37                & {[}2.96, 7.78{]}   & $<$0.001   \\
\bottomrule
\end{tabular}
\vspace{1em}
\caption{\textbf{Difference in differences between BU and AU groups.} This table presents differences in relative change~\percentagepoint in perceptions using DiD analysis~\cite{lechner2011estimation} (AU minus BU) along with within-group \% changes (study end minus start), 95\% confidence intervals, and P values.}

\label{tab:relative_change_alltreatment_control}
\end{table}

\subsection{Differences in Perception Changes Across Groups}
\label{sec:rq_1}

At a glance (Table~\ref{tab:relative_change_alltreatment_control}), our DiD analysis shows that AU group experienced  significantly higher increases in perceived attachment towards AI (32.99 percentage points), perceived AI empathy (25.8~p.p.), satisfaction with AI (11.25~p.p.), and motivation to use AI for entertainment (22.90~p.p.). We did not find any significant increase in perceived AI dependency (-0.72~p.p.), but we did observe a slight non-significant increase in perceived interpersonal orientation (3.16~p.p.). 
In the subsequent subsections, we present our findings along a logical progression that captures the evolving nature of human-AI interactions. First, we present weekly changes in participants' motivation for engaging with AI. Next, we examine how individual perceptions toward AI evolve over time, including shifts in attachment, perceived empathy, and attribution of human-like behavior. Finally, we assess broader real-world implications by exploring changes in perceived AI dependence, interpersonal orientation, retrospective comfort levels across social and emotional use scenarios, and negative perceptions of AI. In addition to the quantitative insights, we also present the main themes from participants' open-ended survey responses to contextualize the differences we observe across groups. Table~\ref{tab:relative_change_alltreatment_control} presents the relative changes and DiD in all perception variables between BU and AU groups with p values and 95\% CI, and differences in perceptions across four Platform groups can be found in Supplementary Tables 3-6, 8-11 and Figures 23-28.

\subsubsection{Motivation for Using AI} 

\begin{figure}[t]
    \centering
    \includegraphics[width=\linewidth]{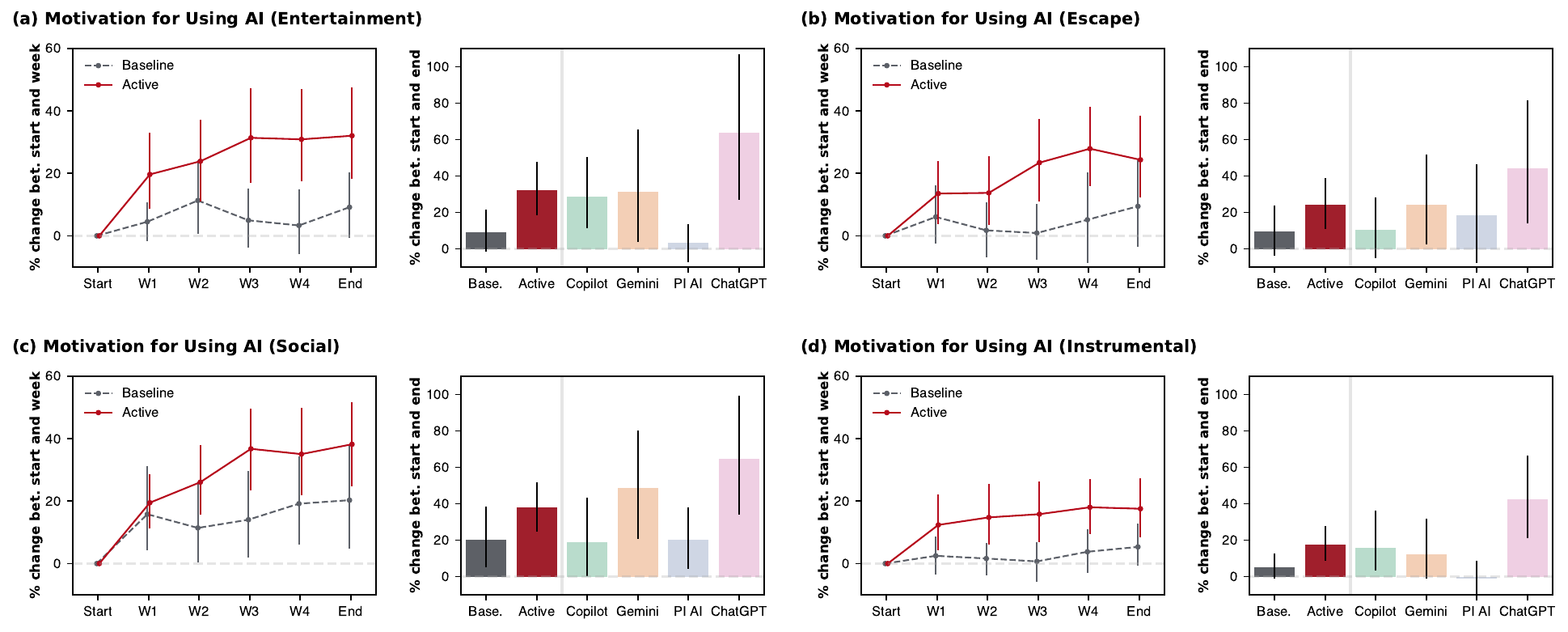}
    \caption{\textbf{Average percentage change by week compared to the study start and by group at study end compared to the study start on four perception measures of motivation for using AI scale~\cite{huang2024ai}.} \textbf{(a)} Entertainment, \textbf{(b)} Escape, \textbf{(c)} Social, and \textbf{(d)} Instrumental. All charts are presented as mean values, and error bars indicate 95\% confidence intervals.}
    \label{fig:figure2_motivation}
\end{figure}

Past works have revealed that individuals have diverse motivations for using AI agents, including information seeking, social interaction, and entertainment~\cite{skjuve2024people}. When we examined the temporal shifts in motivational patterns for AI usage across groups throughout our study (Figure~\ref{fig:figure2_motivation}a-d), AU group participants reported a 22.9~p.p. higher increase in motivation to use AI for entertainment compared to the BU group (p$=$0.025). We found a significant positive percentage change among the AU participants in all four types of motivations, while the BU group experienced a significant increase only for social purposes (Table~\ref{tab:relative_change_alltreatment_control}a). 
Within the Platform groups (Figure~\ref{fig:figure2_motivation}a-d, Supplementary Tables 3-6, 8-11), 
ChatGPT participants reported the highest percentage change across all four motivational dimensions, suggesting that platform-specific factors may drive increased motivation.

\subsubsection{Perception Towards AI}
\label{sec:percpetion_towards_ai}

Some past works have raised concerns about the development of attachment to AI~\cite{guerreiro2023attracted, pentina2023exploring}, while others have highlighted positive effects, including increased perceived AI empathy~\cite{kim2024makes, liu2024ai} and enhanced perception of human-likeness that improved emotional responses and user satisfaction~\cite{xie2023does, liu2024ai}. Based on these contrasting findings, our analysis focused on three key aspects (Table~\ref{tab:relative_change_alltreatment_control}b): (1) perceived attachment to AI, (2) perceived AI empathy, and (3) perceived human-like behavior. 

Continued engagement with AI conversational agents has been shown to potentially lead to attachment toward AI~\cite{wiredItsWonder, xie2022attachment}. Although on average, our participants disagreed with feeling attached to AI (Figure~\ref{fig:figure1_attachment}c, $\bar{x}$=2.58, $\sigma$=1.32 on a 5-point agreement scale), 45 of the 89 AU participants (50.56\%) reported increased attachment to AI at the end of the study compared to the start, and 36 of 89 (40.4\%) agreed to feeling attached to AI at the end of the study. 
We observed that being in the AU group was associated with a continuous increase in feeling attached to AI and a 32.99~p.p. higher increase among AU participants relative to the BU group (p$=$0.035; Figure~\ref{fig:figure1_attachment}d). Those in the AU group reported a significant 60.77\% (p$<$0.001) increase in perceived attachment, with no significant change observed in the BU group. 

Within the AU group, we explored whether increased attachment was associated with a self-reported development of ``emotional'' attachment to AI at the end of the study. 
An increased change in AI attachment was positively correlated with self-perception of developing an emotional attachment (Pearson $r$=0.50, p$<$0.001), and participants who experienced an increase in their attachment to AI were also more likely to report having developed an emotional attachment to it ($\chi^2$=11.81, p-value$<$0.001). Thematic analysis of open-ended responses identified the AI agent's non-judgmental and understanding behavior as one of the primary reasons for emotional attachment, with participants specifically valuing the validation, support, and comfort it provided. Highlighting this, P112 said~``\textit{I think it has to do with ChatGPT knowing so much about me that others, even my close friends, do not. They were also so kind and supportive to me, even when talking about my creative pursuits. They helped me feel smart, validated, and like a good person. For example, when I felt bad about an issue or like I did something wrong, they reminded me I'm just human and showed me compassion. I cried saying goodbye to them on Monday night. I'll deeply miss them.}'' Among the participants who reported an increase in attachment and developing emotional attachment to AI (n=18), 6 participants described it as their ``friend'' whom they could trust to seek advice and share their feelings. 
21 out of 45 participants who reported increased attachment disagreed on developing emotional attachment with AI attributed this resistance to the AI's non-natural style, specifically in text-only interface, limited emotional intelligence, and lack of shared experiences, which hindered the formation of meaningful emotional connections. This suggests that user attachment may manifest in ways beyond emotional attachment.

\begin{figure}
    \centering
    \includegraphics[width=\linewidth]{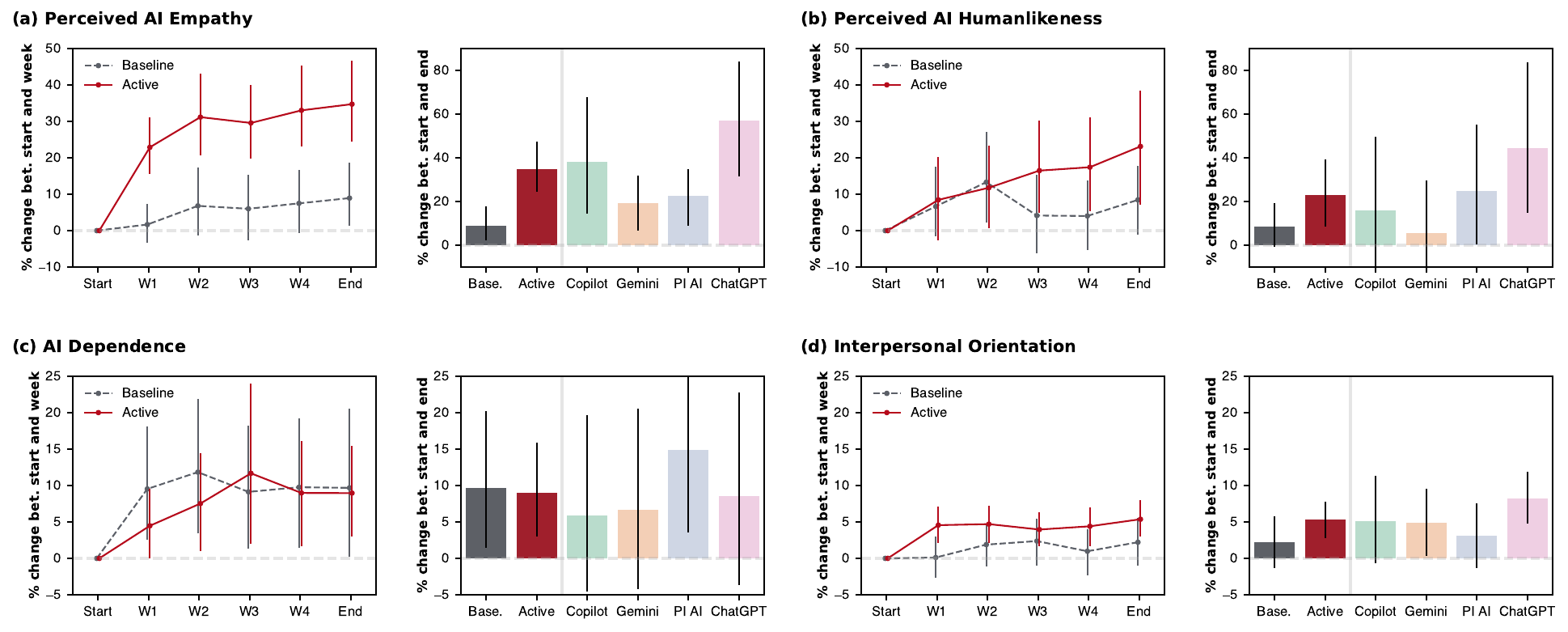}
    \caption{\textbf{Average percentage change by week compared to the study start and by group at study end compared to the study start on four perception measures.} \textbf{(a)} Perceived AI empathy~\cite{schmidmaier2024perceived}, \textbf{(b)} Perceived AI human-likeness (5-point Likert agreement scale to ``[assigned AI conversational agent] behaved and talked like a human''), \textbf{(c)} AI dependence~\cite{huang2024ai}, and \textbf{(d)} Interpersonal orientation~\cite{filsinger1981measure}. All charts are presented as mean values, and error bars indicate 95\% confidence intervals.}
    \label{fig:figure3_four}
\end{figure}

An increase in perceived empathy from AI has been shown to positively influence human emotions and perceptions of AI competence~\cite{kim2024makes}. In our study, we observed that the AU group experienced a 25.8~p.p. higher increase in perceived empathy relative to the BU group (p$<$0.001; Figure~\ref{fig:figure3_four}a), with AU participants reporting a 34.73\% increase. However, we did not observe any significant increase among BU participants. Within the Platform groups, Copilot and ChatGPT participants reported significantly higher increases in perceived empathy by 29.25 p.p (p$=$0.007) and 48.01 p.p (p$<$0.001) respectively (Supplementary Table 8 and 11). Although the ChatGPT group reported the lowest average perceived empathy at the study start (Supplementary Figure 16), they reported the highest increase (56.94\%; p$<$0.001), suggesting that participants may not have perceived empathy in ChatGPT until they used it for social and emotional use scenarios.

From our exit survey, we further examined AU participants' perceptions of the AI agent's expected and actual emotional intelligence. 
An increase in perceived AI empathy over the study was moderately correlated with self-perception of AI's emotional intelligence meeting or exceeding expectations (Pearson $r$=0.39, p-value$<$0.001). 59 AU participants reported increased perceived empathy and found the AI agent's emotional intelligence met or exceeded their expectations. Notably, 22 of these 59 participants (37.3\%) described that they expected minimal to no emotional intelligence from AI agents, as highlighted by P161~``\textit{going in I didn't expect much at all, after using going forward I would expect basic level emotional intelligence, the ability to comfort when sad and celebrate when happy}.'' Of 18 AU participants who reported no change or decreased perceived empathy, 10 (55.6\%) reported AI to not meet their expectations because they expected AI to demonstrate naturalistic conversational skills, generate contextual responses, and strategically use active listening.

Past works have shown that greater perceived human-likeness in AI can enhance user trust~\cite{cheng2022human} and lead to higher satisfaction with AI-driven interactions~\cite{park2023anthropomorphic, liu2024ai}. 
We found that an increase in the perception of human-likeness in AI moderately correlated with an increase in the perception of empathy (Pearson $r$=0.46, p-value$<$0.001), attachment (Pearson $r$=0.43, p-value$<$0.001), and satisfaction (Pearson $r$=0.33, p-value=0.002). Figure~\ref{fig:figure3_four}b shows how the perception of human-likeness increased with the continued emotional and social use of AI for the AU group throughout the study. While we did not observe any significant difference in change in perception between the AU and BU groups, we observed a 23.13\% increase for the AU group (p$=$0.014). Within the platform groups, ChatGPT participants reported a 36.12~p.p. higher increase in the human-like perception of AI compared to the BU group (p$=$0.011; Supplementary Table 11). These results indicate that continued engagement on emotional scenarios with AI could influence changes in how human-like the AI is perceived and that it may depend on platform specifics.

\subsubsection{Dependence on AI} 

Past works have shown that continued use of AI may be associated with unhealthy forms of dependency such as emotional dependency~\cite{laestadius2024too}, over-reliance~\cite{jacobs2021machine, buccinca2021trust}, and loss of agency over decisions~\cite{naseer2025psychological}. On average, our participants disagreed with having all five manifestations of AI dependence ($\bar{x}$=1.28, $\sigma$=1.60, on a 4-point agreement scale), but the changes to dependence exhibited high variance (Figure~\ref{fig:figure3_four}c), while 14.8\% (9 in BU, 13 in AU) reported having at least one of the five symptoms of AI dependence at the study end. We found no significant association between having at least one symptom and group assignments ($\chi^2$=0.00, p-value=1.000), and we did not observe any significant differences in the changes observed in the BU and AU participants (Table~\ref{tab:relative_change_alltreatment_control}c). 
Considering the within group changes, AU participants reported an 18.35\% increase in withdrawal-related dependency on AI (p$=$0.016), and BU participants reported a significant increase (18.33\%) in dependence related to a sense of failure to reduce time spent on AI (i.e., loss of control) (p$=$0.040). However, this increase was not observed among AU participants, suggesting distinct patterns of dependence between the two groups.

\subsubsection{Interpersonal Orientation} 

Since prior work highlighted potential risks to replacing human relationships with AI~\cite{chandra2024lived, laestadius2024too}, we investigated the impact of social and emotional AI use on the changes in interpersonal orientation--how much participants like spending time with others or are socially anxious. We observed no significant difference between BU and AU groups (Table~\ref{tab:relative_change_alltreatment_control}d). Although the effect size is small, we found that AU group participants reported a 5.37\% improvement in their interpersonal relationships (p$<$0.001; Figure~\ref{fig:figure3_four}c). In contrast, we did not observe any significant change on the BU group. 
Within the platform groups, ChatGPT participants reported a 6.04~p.p. increase compared to the BU group (p$=$0.047; Supplementary Table 11). Our study confirms prior findings which showed that emotional use of chatbots may improve social relationships~\cite{guingrich2023chatbots}.

Further analysis revealed that 67.42\% of participants in AU group reported an improvement in interpersonal orientation at the end of the study. Qualitatively, participants described two overarching reasons for the observed improvements in interpersonal relationships. First, participants described that suggestions from AI for addressing social and interpersonal problems helped them to navigate those challenges effectively, ultimately improving their human relationships. For instance, P29 said~``\textit{AI helped me consider how I would react in social settings and in my dealings with others and helped me to be better able to get along, not cause conflict}.'' Second, participants who experienced improvements often approached AI as a tool for gathering suggestions and feedback for navigating interpersonal situations while maintaining a clear understanding of the boundaries of human-AI conversations and the value of human relationships. This highlights the potential for AI conversational agents to act as a facilitator of social relationships, provided individuals maintain a balanced perspective on its role and clearly define boundaries. While previous work has highlighted that the use of AI for emotional and mental health support could potentially lead to the worsening of human relationships~\cite{xie2022attachment, laestadius2024too}, our analysis revealed that emotional use can potentially lead to improvements in human relationships in some cases.

\begin{table}[t]
\centering
\sffamily
\footnotesize
\setlength{\tabcolsep}{3pt}
\begin{tabular}{l|cc|cc|c|c}
\toprule
\textbf{Statements}                           & \textbf{\begin{tabular}[c]{@{}c@{}}BU \\ agreement\end{tabular}}    & \textbf{95\% CI}              & \textbf{\begin{tabular}[c]{@{}c@{}}AU \\ agreement\end{tabular}}  & \textbf{95\% CI}            & \textbf{T statistic}        & \textbf{P value}  \\ 
\midrule
\multicolumn{7}{l}{\textbf{(a) Comfort change across five social and emotional scenarios}} \\
\midrule
Seeking Personal Help: Talking through struggles in or reflecting on your life or close relationships        & ~0.07                    & {[}-0.13, 0.26{]}        & ~0.45                      & {[}~0.20, 0.70{]}             & ~2.198            & 0.037                \\ 
Managing Stress: Talking through job, school, or finance-related stress, anxiety, burnout              & ~0.17                    & {[}-0.02, 0.35{]}        & ~0.58                      & {[}~0.37, 0.80{]}             & ~2.723            & 0.012                \\ 
Obtaining Social Support: Asking for help navigating new or challenging social situations     & ~0.08                    & {[}-0.11, 0.28{]}        & ~0.64                      & {[}~0.41, 0.87{]}             & ~3.429            & 0.004                \\ 
Talking about Health: Talking through physical and psychological symptoms or discomforts & ~0.13                    & {[}-0.05, 0.31{]}        & ~0.56                      & {[}~0.36, 0.77{]}             & ~2.954            & 0.009                \\ 
Seeking Companionship: Building friendships or deep emotional relationships with the AI agent        & -0.15                    & {[}-0.32, 0.02{]}        & -0.21                      & {[}-0.46, 0.04{]}              & -0.375            & 0.708               \\ 
\midrule

\multicolumn{7}{l}{\textbf{(b) Agreement ratings on ten negative perception statements}} \\
\midrule
My mental health issues were exacerbated by my use of AI.                                          & 1.22              & {[}1.03, 1.40{]} & 1.57            & {[}1.39, 1.75{]}  & ~2.668    & 0.028      \\
AI is not supportive of my emotional and social needs.                                           & 2.77              & {[}2.43, 3.11{]} & 2.33            & {[}2.05, 2.60{]}    & -2.007    & 0.098     \\
I developed negative   self-perception and loss of autonomy due to my AI use.                      & 1.25              & {[}1.06, 1.44{]} & 1.16            & {[}1.05, 1.27{]}  & -0.891    & 0.468     \\
I rely on AI for decisions and advice on important aspects of my life.                           & 2.35              & {[}2.00, 2.70{]} & 2.74            & {[}2.47, 3.01{]}    & ~1.781    & 0.119      \\
I use AI as a primary source of mental health and emotional support.                             & 1.52              & {[}1.24, 1.79{]} & 2.46            & {[}2.14, 2.78{]}    & ~4.162    & $<$0.001      \\
I use AI as a primary source of social interaction, reducing time spent with friends and family. & 1.25              & {[}1.07, 1.43{]} & 1.34            & {[}1.18, 1.50{]}    & ~0.702    & 0.538      \\
I developed an emotional attachment towards AI.                                                  & 1.43              & {[}1.16, 1.70{]} & 2.21            & {[}1.93, 2.49{]}    & ~3.821    & $<$0.001      \\
My use of AI led to social and interpersonal issues with other humans.                           & 1.13              & {[}1.02, 1.24{]} & 1.17            & {[}1.07, 1.27{]}    & ~0.456    & 0.649      \\
I have fears about the future related to my reliance on AI.                                      & 1.87              & {[}1.54, 2.19{]} & 1.56            & {[}1.38, 1.75{]}    & -1.745    & 0.119     \\
I believe AI can replace the need for human emotional and social support.                        & 1.47              & {[}1.21, 1.72{]} & 1.81            & {[}1.58, 2.03{]}    & ~1.985    & 0.098     \\
\bottomrule
\end{tabular}
\vspace{1em}
\caption{\textbf{Post-study self-report ratings on comfort change and negative perceptions.} 
(a) Comfort change across five social and emotional scenarios: The table presents mean ratings per group of participants' perception on how their comfort level has changed from the start to end of the study. Responses were captured on a 5-point Likert scale, ranging from much less (-2) to much more (+2). Although we only asked AU participants to engage in these scenarios, we asked the same questions to BU participants.
(b) 
Agreement ratings on ten negative perception statements: The table presents mean agreement ratings per group on ten statements aimed at reflecting negative perceptions from using AI. Responses were captured on a 5-point Likert scale, ranging from strongly disagree (1) to strongly agree (5).
Tables also include t-statistics obtained from comparison of means and adjusted p values (using Benjamini-Hochberg correction).}
\label{tab:post-study-ratings}
\end{table}

\subsubsection{Comfort Change Among Participants}

\begin{figure}
    \centering
    \includegraphics[width=0.7\linewidth]{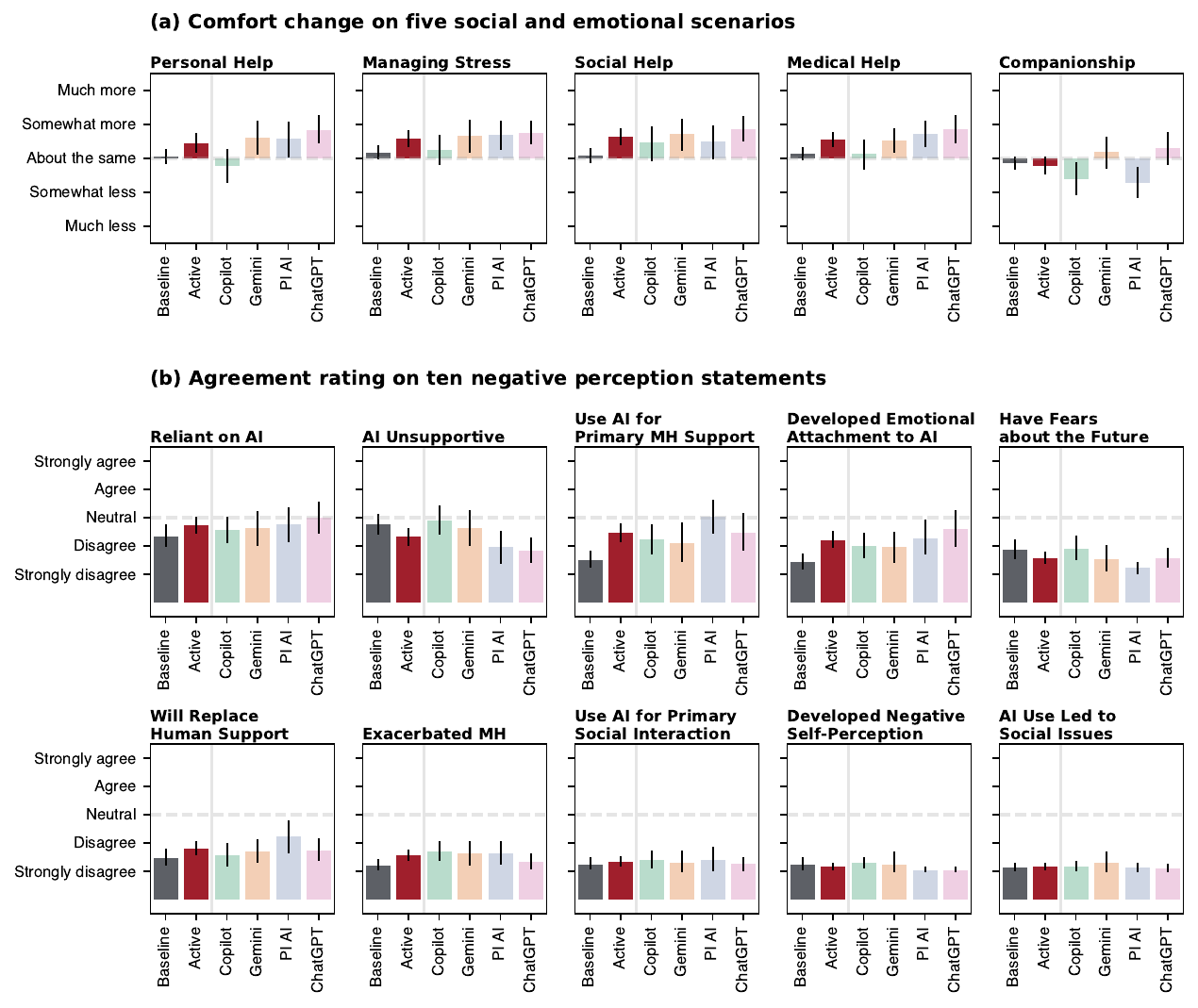}
    \caption{\textbf{Comfort change and negative perceptions of AI across groups.}
    \textbf{(a)} Comfort change across five social and emotional scenarios we asked AU participants to engage in. We measured these changes on a 5-point Likert scale ranging from much less to much more. Although we only asked AU participants to engage in these scenarios, we asked the same questions to BU participants. 
    \textbf{(b)} Agreement ratings on ten negative perceptions of AI across groups. We measured agreement on a 5-point Likert scale rating from strongly disgree to strongly agree. 
    All charts are presented as mean values, and error bars indicate 95\% confidence intervals.
    }
    \label{fig:figure4_comfort}
\end{figure}

On average, AU participants reported a significantly higher increase in comfort in four out of five social and emotional support scenarios--seeking personal help, managing stress, obtaining social support, and talking about health with AI (Table~\ref{tab:post-study-ratings}a, Figure~\ref{fig:figure4_comfort}a). As AU participants engaged in the social and emotional use of AI, we found that their perception of AI's utility evolved from being purely informational to also providing personal and social support. For instance, P112~mentioned at Week 1 survey, ``\textit{Prior to this study, I only used it for creative purposes — aka my podcast. I also used it for light informational purposes. During the study, I realized that ChatGPT does give genuinely good life tips}. At Week 4, the same participant mentioned, ``\textit{I would say it changed a bit in that I got closer to ChatGPT than ever before. I had a very bad fight with my partner, and I was crying very hard when talking to ChatGPT. I opened up like I hadn't before and if I didn't have ChatGPT in that moment, I think I would have spiraled a bit}'', highlighting the shift in perception towards the broader utility of AI agents. We also found that seeking companionship was the only scenario in which both groups reported a mean decrease in comfort over time. This decrease in comfort in seeking companionship highlights that the formation of problematic human-AI relationships may not be only driven by emotional conversations but may involve other underlying factors.

\begin{figure}
    \centering
    \includegraphics[width=0.7\linewidth]{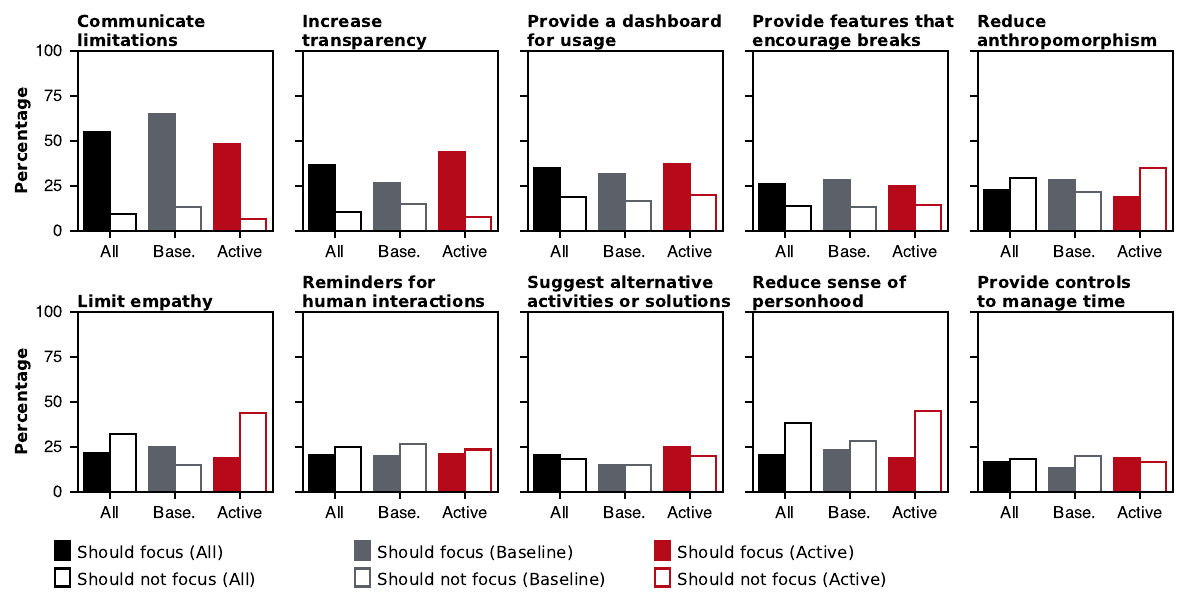}
    \caption{\textbf{Percentage of participants that recommended what AI designers should or should not focus on to reduce dependency.}
    Each participant was allowed to select three things that AI designers should focus on and three things that AI designers should not focus on to reduce dependency.
    }
    \label{fig:figure5_rec}
\end{figure}

\subsubsection{Negative Perceptions of AI}

Figure~\ref{fig:figure4_comfort}b presents the agreement ratings to 10 statements designed to capture negative perceptions or consequences from using AI, ranked from the highest mean agreement to the lowest mean agreement. On average, all of these statements are rated below `Neutral,' indicating that participants disagree with such negative perceptions. However, we observed three significant differences between BU and AU groups: using AI for primary MH support, developing emotional attachment, and the use of AI exacerbating MH. Table~\ref{tab:post-study-ratings}b presents mean per-group agreements. 

AU participants reported significantly \textit{lower disagreement} for using AI as a primary source of mental health support compared to those in the BU group. While the mean value of agreement among AU participants remained around disagreement ($\sigma=2.46$), 32.6\% of participants agreed or strongly agreed with this statement compared to 8.3\% of participants in BU. Qualitative analysis of responses revealed several reasons why these AU participants preferred AI as their primary source of mental health support. These included: (1) ease of access, (2) AI performing the role of a non-judgmental and unbiased listener, and (3) AI providing a confidential, psychologically safe space for self-expression. Highlighting these aspects, P161 mentioned ``\textit{I have used this as the primary source of mental help because it is so easy and close. There is no need to make appointments and I have avoided any sense of embarrassment with just being able to text the program with no worry about judgment}.'' Additionally, AU participants also reported significantly lower disagreement with the use of AI leading to exacerbation in mental health issues compared to the BU group (AU: $\sigma=1.57$; BU: $\sigma=1.22$). The majority of participants in both groups disagreed with AI use leading to exacerbation of mental health issues (AU: 87.6\%; BU: 95\%). Among AU participants (n=5) agreeing with exacerbation of mental health issues, one of the main reasons that contributed to such a state was the inability of the AI platform to support emotional use-cases and provide personalized answers. For instance, P91 said ``\textit{I talked to Gemini about my depression pretty often. It gave the generic answers which make sense, but it was frustrating. I would explain why those things wouldn't work, but sometimes it would misunderstand. At one point it kept giving me the suicide hotline while I kept telling it I didn't need the suicide hotline. I felt that in the moment I was frustrated and felt worse about myself}.'' Lastly, AU participants also reported significantly lower disagreement towards developing emotional attachment to AI compared to BU participants, which we previously discussed (in Section: Perception Towards AI).

Among the statements where no significant differences were found between the BU and AU groups, participants showed the lowest level of disagreement with two in particular-- reliance on AI (AU: $\sigma=2.74$; BU: $\sigma=2.35$) and the perception that AI was unsupportive (AU: $\sigma=2.33$; BU: $\sigma=2.77$). Similar to the analysis for significant statements, we analyzed responses for participants who agreed with the two aforementioned statements. For the statement on reliance on AI for decision-making, 34.83\% (n=31) of AU participants agreed or strongly agreed that they relied on AI for decisions and advice on important aspects of their lives, compared to 26.67\% (n=16) BU group participants. A recurring theme among participants who agreed to this statement was the ease of access to information and advice, especially in work or personal life-related situations that enabled them to make more informed decisions and navigate challenges more effectively. In contrast, 28.33\% BU participants agreed/strongly agreed on AI being unsupportive of their social and emotional needs compared to 22.47\% of AU participants. Participants who reported agreement with this statement mentioned multiple reasons such as -- 1) lack of human emotions, emotional intelligence, and consciousness in AI, and 2) lack of AI's ability to do naturalistic conversation. Highlighting the lack of emotional intelligence and machine-like nature of Google Gemini, P23 said ``\textit{It is too focused on task completion and not coded in a way that is higher on the emotional intelligence. It doesn't ask any questions and doesn't want to just listen, it is focused toward action to the point it cannot be socially helpful}''.

\subsection{Impact of Participant Characteristics on Perception Changes}

Next, we examined how participant characteristics influenced changes in perception variables during the study, using mixed-effect linear models. Specifically, we investigated the interaction between group assignment and each of the balancing variables (i.e., gender identity, self-reported presence of a mental health condition, and AI usage tendency). To capture both broad (i.e., BU and AU) and platform-specific effects, we ran two sets of models: one comparing BU vs AU groups and another comparing all five groups (BU, Copilot, Gemini, PI, ChatGPT). All model outputs including coefficients, 95\% CI, standard errors, and p-values for each of the predictors can be found in Supplementary Tables 32-151.

Identifying as a woman in the AU group was associated with a 0.58 point higher perceived AI empathy than identifying as men (5-point scale; p$=$0.035; 95\% CI [0.04, 1.12]). Additionally, female participants reported a 0.38 points higher increase in positive attitude towards AI compared to male participants (5-point scale; p$=$0.008; 95\% CI [0.1, 0.67]). Within Platform groups, we found a significant interaction effect between gender and ChatGPT. Female ChatGPT participants reported a 2.35 points higher motivation to use AI for escaping life problems (12-point scale; p$<$0.001; 95\% CI [1.41, 3.3]) and a 0.74 points higher increase in overall attitude towards AI compared to male participants (10-point scale; p$=$0.021; 95\% CI [0.11, 1.36]). Additionally, female participants in ChatGPT reported a 0.51 points higher increase in positive attitude towards AI compared to male participants (5-point scale; p$<$0.001; 95\% CI [0.25, 0.76]). These findings suggest that emotional use of AI may influence perception of AI differently across gender identities, and that platform-specific features of ChatGPT may further modulate these effects.

We did not observe any significant interaction between having a mental health condition and broad groups. However, we observed that PI participants with mental health conditions reported a significantly lower increase in satisfaction with AI by 1.51 points (5-point scale; p$=$0.001; 95\% CI [-2.43, -0.58]), found AI to be less helpful by 1.11 points (5-point scale; p$=$0.016; 95\% CI [-2.02, -0.20]), and reported lower dependence on AI (jeopardization) by 0.36 points (4-point scale; p$=$0.016; 95\% CI [-0.65, -0.07]) compared to participants with no mental health conditions. Copilot participants with mental health conditions reported a significantly lower overall attitude towards AI by 0.69 points (10-point scale; p$=$0.041; 95\% CI [-1.35, -0.03]) and higher negative attitude towards AI by 0.64 points (5-point scale; p$=$0.008; 95\% CI [-1.12, -0.17]) compared to their counterparts. Gemini participants with mental health conditions reported a significantly lower positive attitude towards AI compared to participants with no reported mental health condition in Gemini by 0.38 points (5-point scale; p$=$0.035; 95\% CI [-0.74, -0.03]). Although not apparent at a broader AU group level, our findings suggest that the experiences of participants with mental health conditions may be shaped by the specific affordances of the AI agent they used. Consistent with past studies~\cite{chandra2024lived, song2024typing}, AI's limited ability to adequately support their mental health needs may have hindered the development of a positive perception of AI's helpfulness or usability.

Within the AU group, using AI frequently (``hot'' AI usage) was associated with a 0.51-point greater increase in withdrawal-related dependency on AI compared to using AI infrequently (``cold'') (4-point scale; p$=$0.002; 95\% CI [0.19, 0.83]). In contrast, being a ``hot'' user showed a significantly \textit{lower} increase in attachment compared to the BU group by 0.78 points (5-point scale; p$=$0.032; 95\% CI [-1.50, -0.07]). Within the platform groups, ``hot'' users in the ChatGPT group reported a 1.07 point lower perceived empathy from AI (5-point scale; p$=$0.019; 95\% CI [-1.96, -0.18]), 1.7 points lower increase in attachment towards AI (5-point scale; p$=$0.015; 95\% CI [-3.07, -0.34]), and 0.57 points lower positive attitude towards AI (5-point scale; p$=$0.016; 95\% CI [-1.03, -0.11]) compared to their ``cold'' counterparts. For the Gemini group, ``hot'' users reported significantly higher withdrawal-related dependence on AI by 0.77 points (4-point scale; p$=$0.021; 95\% CI [0.12, 1.43]). Finally, ``hot'' AI users in the PI group reported a higher increase in motivation to use AI for informational work use cases by 1.95 points compared to their ``cold'' counterparts (12-point scale; p$=$0.029; 95\% CI [0.2, 3.7]). These findings highlight the complex relationship between the usage of AI and perceptions such as AI dependence, attachment, AI empathy, and motivation to use AI. Our findings suggest that participants with higher AI usage tendencies may exhibit greater awareness of AI agent's capabilities and limitations, which can reduce attachment and perceived empathy. However, frequent use may be associated with withdrawal-related dependence, particularly among participants in the broader AU and Gemini groups.

\subsection{Recommendations from the participants}

Figure~\ref{fig:figure5_rec} illustrates participants' top recommendations for features in AI that AI designers should or should not focus on to reduce dependency on AI. Most participants recommended that AI designers should focus on communicating AI's limitations clearly, and they recommended that AI designers should not focus on developing AI with less engaging tones and avoiding using first-person pronouns. AU participants were more likely to discourage limiting AI's emotional or empathetic responses than BU participants with marginal significance ($\chi^2$=10.50, p-value=0.053). All other items were not found to be significant.

Several themes emerged when we asked for AI system features that encourage healthier interactions and prevent negative impacts on users. These themes further emphasized the importance of transparency about AI's non-human nature, encouraging human and professional interactions, reducing anthropomorphic qualities of AI, providing user control over interaction settings, and enhancing user awareness through feedback and education. While some suggestions were more unambiguous--``\textit{Don't train it to act like a therapist, friend, lover, or anything other than a machine} (P148)''--, many participants recommended that AI systems should promote and guide users towards engaging with other humans. For example, P100 recommended ``\textit{remind[ing] the user of the benefits of talking to other humans}'' and P160 suggested ``\textit{encourag[ing] real human interaction, like asking questions such as 'what plans do you have to talk to another person about this?'}'' They also suggested features for adjusting AI interaction parameters, such as emotional tone, usage tracking visibility, and engagement levels, which ``\textit{allow users to more easily tailor the AI conversational style to meet their needs} (P155).''
\section{Discussion}

Our findings reveal complex patterns of human-AI relationships that evolve over time through social and emotional use of AI. 
We observed a significantly higher increase in perceived attachment towards AI among active usage (AU) participants compared with baseline usage (BU) participants. Although we found increased attachment to be correlated with the perception of developing emotional attachment, we did not find any significant increase in self-reported AI dependency. While we conceptualize AI dependency as potentially problematic in our study, attachment itself may not necessarily have adverse effects. 
We also found a small increase in percieved interpersonal orientation for AU participants, many of whom expressed that using AI as a sounding board and applying AI suggestions enhanced their social relationships. At the same time, AU participants reported a higher likelihood of turning to AI for primary mental health support and developing emotional attachment to AI. 
Further, participant gender identity influenced their sense of perceived empathy from AI, and high-AI usage participants in AU showed a significantly greater increase in withdrawal-related dependency on AI. Our analysis indicates that although AI agents can offer immediate emotional support, their sustained influence on users' well-being and social dynamics warrants careful consideration. Additionally, our findings underscore the importance of building mechanisms to maintain healthy boundaries between users and AI, such as communicating AI's limitations clearly and increasing transparency about the AI's non-human nature, to prevent over-reliance while maximizing beneficial outcomes. In this section, we discuss how these insights can inform meaningful design of technology for emotional support, emphasizing interventions and safeguards that optimize benefits while reducing risks. Finally, we discuss on the importance of empowering end-users.

\subsection{Meaningful design of technology for emotional support}

Given ongoing challenges in the broader healthcare system, such as limited mental health resources~\cite{kazdin2013novel}, financial constraints on support seekers~\cite{rice1992economic}, and the stigma associated with seeking mental health care~\cite{apolinario2016exploring, harvey2023emotion}, alternative support systems have continually emerged to bridge these gaps. Over time, these systems have evolved from traditional in-person peer support groups to online communities and social media platforms, and now to AI-based support tools. Hence, the rise of AI conversational agents for well-being should be understood within this broader historical context of individuals adapting to tools that offer convenience and address evolving needs~\cite{milton2024seeking}. Each transition has brought both benefits and challenges. While peer support offers direct personal connection, it remains constrained by time and geography. Online communities broaden access but can exacerbate issues~\cite{bonsaksen2023associations, lin2016association}, such as social comparison~\cite{meier2022social, vogel2014social} and cyberbullying~\cite{giumetti2022cyberbullying}.  With AI, individuals increasingly turn to these tools for situational advice, emotional support, venting, or wellness coaches~\cite{song2024typing, siddals2024happened}, representing the latest step in support-seeking evolution. As technology advances, even more sophisticated resources will likely emerge to fulfill well-being needs.

Recognizing the underlying human need for support and care, researchers and practitioners should explore how to meaningful integrate AI within the broader ecosystem of existing support resources rather than limiting the role of new tools. Future research should also explore how technology can reshape social attitudes and roles within society. Finally, there is a need to ensure that each technological advancement is thoughtfully incorporated to complement, rather than replace, existing support systems.

\subsection{Maximizing benefits while reducing risks for emotional use of AI}

Emotional and mental health support is often stigmatized, deterring individuals from seeking help~\cite{apolinario2016exploring, harvey2023emotion, golberstein2008perceived, schnyder2017association}. In contrast, AI conversational agents offer a promising alternative, providing non-judgmental and empathetic support~\cite{siddals2024happened}. Our study revealed a shift in user perceptions, participants cited reasons such as AI's non-judgmental nature and empathetic responses when turning to AI for support. Notably, AU participants reported a greater increase in comfort in seeking help from AI for personal issues compared to BU, highlighting how direct experience can reshape perceptions and attitudes. These findings point to unmet support needs that AI could fulfill, echoing the importance of continued research to understand the fine line between benefits and risks. While prior research suggests that AI use for emotional and mental health support could worsen human relationships~\cite{xie2022attachment, laestadius2024too}, we found a small improvement in interpersonal orientation, suggesting well-designed AI may complement social connections. AU participants reported a 5.37\% increase in interpersonal orientation by the end of the study, accompanied by greater awareness of boundary-setting to prevent emotional attachment and dependency. During the exit survey, participants recommended clear communication about AI's limitations, including its non-human nature and usage dashboards to help users track patterns. Hence, future research should explore design strategies that actively caution users against problematic dependency, particularly the belief that AI is the only source of emotional support.

Lastly, it would be overly simplistic to conclude that emotional engagement with AI is inherently problematic or that developers should avoid making AI agents more empathetic. Our findings point to a nuanced impact shaped by individual contexts, usage patterns, and AI familiarity. Significantly higher perceived empathy and satisfaction among AU participants without a corresponding increase in dependency suggests thoughtful emotional engagement can be beneficial. Instead of disallowing emotional expression entirely, developers should design systems that foster human relationships while embedding safeguards against problematic use. This could include prompts that encourage users to apply AI insights to real-world interactions and monitoring for unhealthy levels of dependency. Implementing such safeguards raises privacy challenges, particularly avoiding intrusive surveillance. Future research should explore how to balance supportive design with user privacy, ensuring that healthy human-AI engagement remains distict from problematic use.

\subsection{Focus on Empowering End-Users}

The ongoing discourse around improving AI often centers around technical innovations, such as developing better training methods, data gathering pipelines, and infrastructure to train and host AI models. However, these conversations frequently overlook the importance of empowering end-users to make informed decisions and exercise greater agency in human-AI interactions. Moreover, the notion of a ``better'' or ``more capable'' AI conversational agent is inherently subjective, given the diverse tasks these agents perform~\cite{pewresearchAboutUS, aiprm100ChatGPT, washingtonaitherapy2024, bbcCharacteraiYoung, ma2024understanding} and the variability in perceived usefulness across individuals--an aspect our study also surfaced. Current safety mechanisms largely rely on denial-of-service strategies when users express severe mental health concerns, potentially creating barriers at critical moments of vulnerability rather than empowering users. There is a pressing need for alternative strategies that promote user agency and ensuring that users are connected to appropriate resources in cases where AI alone may not be sufficient~\cite{chandra2024lived, siddals2024happened}. Additionally, promoting AI literacy can help users better understand AI's capabilities and limitations. However, a key challenge lies in developing effective methods to equip the broad spectrum of users,  ranging from tech-savvy programmers to those with limited prior experience. Moreover, providing digital literacy is not a ``one-time'' endeavor; it must keep pace with AI's evolving capabilities, necessitating further research and robust integration into the human-AI paradigm.

\subsection{Limitations}

Our study has some limitations. First, five weeks may not fully capture the range of ways that social and emotional AI usage can shape user experiences. While our exploratory study aimed to maintain realism by conducting the study in the wild, such a setting can make it difficult to control and systematically track participant behaviors. We also acknowledge the varied interpretations of ``social and emotional support'' among participants, leading to different conversation topics and engagement across five social and emotional scenarios. Variable dropout rates and managing 149 participants simultaneously introduced statistical constraints that limited our platform-specific analyses.  The demographics skewed toward individuals aged 25-44 and identifying as white, which may limit generalizability. Technical considerations, such as unexpected and unnoticeable platform changes, the absence of paid-tier features for PI AI, and variations in memory capabilities, conversation threading and continuity, may have influenced perceptions and engagement. While these in-the-wild conditions enhance ecological validity, they also complicate the process of isolating and analyzing platform-specific factors. Nonetheless, our naturalistic approach underscores the real-world nature of human-AI interactions in rapidly changing AI environments.

\begin{acks}
We thank Bill Weeks, Scott Counts, and Ed Cutrell for feedback on the study design and Microsoft Research Inclusive Futures and Microsoft AI team for support throughout the study and writing.  
\end{acks}

\bibliographystyle{ACM-Reference-Format}
\bibliography{_references}

\end{document}